\documentclass[usenatbib]{mn2e}

\usepackage{times}
\usepackage{graphicx}
\usepackage{mathrsfs}

\title[Detecting starlight scattered by transiting hot Jupiters]{A comparison of spectroscopic methods for detecting starlight scattered by transiting  hot Jupiters, with application to Subaru data for HD~209458b and HD~189733b}
\author[Langford {\it et al.}]{Sally~V.~Langford,$^{1,2}$ J. Stuart B. Wyithe,$^1$ Edwin L. Turner,$^{2,3}$ Edward B. Jenkins,$^2$ 
\newauthor 
Norio Narita,$^4$ Xin Liu,$^2$ Yasushi Suto$^{2,5,6}$ and Toru Yamada$^7$ \\
$^{1}$School of Physics, University of Melbourne, Parkville, Victoria 3010, Australia\\
$^{2}$Princeton University Observatory, Princeton NJ 08540, USA\\
$^{3}$Institute for the Physics and Mathematics of the Universe, University of Tokyo, Kashiwa 277-8568, Japan\\
$^{4}$National Astronomical Observatory of Japan, 2-21-1 Osawa, Mitaka, Tokyo, 181-8588, Japan\\
$^{5}$Department of Physics, The University of Tokyo, Tokyo 113-0033, Japan\\
$^{6}$Research Center for the Early Universe, School of Science, University of Tokyo, Tokyo 113-0033, Japan\\
$^{7}$Astronomical Institute, Tohoku University, Sendai 980-8578, Japan\\}
\begin{document}

\label{firstpage}

\maketitle

\begin{abstract}

The measurement of the light scattered from extrasolar planets 
informs atmospheric and formation models.
With the discovery of many hot Jupiter planets orbiting nearby
stars, 
this motivates 
the development of robust methods of characterisation from follow up
observations.
In this paper we discuss two methods for determining the planetary albedo in transiting systems. 
First, the most widely used method for measuring the light scattered by hot Jupiters \citep{Cameron1999} is investigated for application for  typical \'{e}chelle spectra of a transiting planet system, showing that a detection requires high signal-to-noise ratio data of bright planets. 
Secondly a new Fourier analysis method is also presented, which is model-independent and utilises the benefits of the reduced number of unknown parameters in transiting systems. This approach involves solving for the planet and stellar spectra in Fourier space by least-squares.
The sensitivities of the methods are determined via Monte Carlo simulations for a range of planet-to-star fluxes. We find the Fourier analysis method to be better suited to the ideal case of typical observations of a well constrained transiting system than the \citet{Cameron1999}  method.
To guide future observations of transiting planets with ground-based capabilities, the expected sensitivity to the planet-to-star flux fraction is quantified as a function of   signal-to-noise ratio and wavelength range.
We apply the Fourier analysis method for extracting the light scattered by transiting hot Jupiters from high resolution spectra to \'{e}chelle spectra of HD~209458 and HD~189733.
Unfortunately we are unable to improve on the previous upper limit of the planet-to-star flux for HD~209458b set by space-based observations. 
A $1\sigma$ upper limit on the planet-to-star flux of HD~189733b is measured in the wavelength range of 558.83--599.56~nm yielding $\epsilon<4.5\times10^{-4}$. This limit is not sufficiently strong to constrain models. Improvement in the measurement of the upper limit of the planet-to-star flux of this system, with ground-based capabilities, requires data with a higher   signal-to-noise ratio, and increased stability of the telescope.

\end{abstract}

\section{Introduction}

More than 500 planets have been discovered outside the Solar System in the past two decades.  
Among these planets is a class of objects known as hot Jupiters.  
These are roughly Jupiter-mass objects that orbit their host stars within 0.05 AU \citep{Seager2000}, but not close enough to be considered very hot Jupiters, where thermal emission dominates the visible scattered light.  
It has been predicted that hot Jupiter planets will allow the starlight scattered by their upper atmospheres to be directly detected in the optical, with an expected flux of less than $10^{-4}$ times the direct stellar flux \citep{Cameron1999, Charbonneau1999}. 

With increasing numbers of hot Jupiter planets currently being discovered, follow up observations become very important for characterisation that requires longer observation times, and for measuring the accuracy of the orbital parameters. 
The space-based CoRoT and Kepler missions, which are currently monitoring many thousands of nearby stars, will yield a wealth of new transiting planets. 
These planets will also be followed up with radial velocity measurements to constrain the orbital parameters. 
In order to fully realise the scientific potential of these systems, robust methods of analysis of the starlight scattered by the transiting hot Jupiter planets are required, as well as an understanding of their application to various systems. 

There has been no definite detection of the broadband unpolarised visible light scattered by a hot Jupiter. 
Extrasolar planets have only recently been directly imaged in the optical. 
These planets orbit their parent stars at $\sim$100~AU \citep{Kalas2008}. 
In addition, visible polarised scattered light has been detected from
the extended atmosphere of the transiting planet HD~189733b
\citep{Berdyugina2008,Berdyugina2011}. 
However, \citet{Wiktorowicz2009} was unable to reproduce this result. 
Measuring the light scattered by a hot Jupiter planet is challenging due to the small angular separation, and the small planet-to-star flux ratio, which can be of the order of the systematic noise of ground-based observations. 
Extracting the planet signal from the combined noisy stellar and scattered flux therefore generally requires making assumptions about the system.
Upper limits on the visible light scattered by a hot Jupiter planet have been measured with ground-based and space-based capabilities \citep{Charbonneau1999, Cameron2002, Leigh2003a, Rowe2008,Kipping2011, Alonso2009a, Alonso2009b, Rogers2009, Christiansen2010, Sing2009}.

The light scattered by an extrasolar planet will depend on the composition of the atmosphere as well as the orbital parameters, allowing the testing of atmospheric models \citep{Marley1999,  Green2003}.  
Models predict a range of planet-to-star fluxes for different classes of planetary atmospheres, which can guide the search for thermal emission and reflected light from hot Jupiter planets \citep{Sudarsky2000,Sudarsky2003}. 
The non-detections of scattered light and resulting upper limits on the albedos of hot Jupiter planets aids the development of theories of extrasolar planet atmospheres. 
The lack of a detection of starlight scattered by the planet HD~209458b suggests that the atmosphere is much less reflective than previously expected, and lacks highly scattering clouds \citep{Seager2000, Rowe2006,Burrows2008}. 
Increasingly sophisticated models are attempting to explain the inflated radius and absence of a verifiable detection of light scattered by this hot Jupiter planet  \citep{Hood2008, Burrows2008}. 
Further investigation and observation of known hot Jupiter planets will continue to guide the models, and potentially yield a definite detection of scattered broadband unpolarised visible light. 

\subsection{Current Methods}

Prior to the discovery of transiting extrasolar planets, two methods of measuring the planet-to-star flux of hot Jupiter planets were concurrently developed by \citet{Charbonneau1999} and \citet{Cameron1999}.
Both methods require modelling the direct stellar contribution, which is then removed, leaving the planet signal buried in noise. 
They model the system over a range of orbital phases, inclinations and velocities in order to find the signature of the planet's scattered light, and the likelihood of the best fit orbital parameters. 
Both methods been important in beginning to constrain models of planetary atmospheres \citep{Sudarsky2000, Sudarsky2003}.

An upper limit on the planet-to-star flux of the extrasolar planet $\tau$~Bo\"{o}tis b was measured to be $\leq 5\times 10^{-5}$ times the direct stellar flux at the 99~per cent confidence level \citep{Charbonneau1999}.  
This measurement corresponds to a wavelength range of 465.8--498.7~nm, and assumes an inclination of $i \geq70^{o}$.   
The result was limited by the photon noise of the data, not systematic effects. 

In disagreement with this, \citet{Cameron1999} measured a best fit to the planet-to-star flux of the same system to be $(7.5\pm3) \times~10^{-5}$ at the $97.8$~per cent confidence level. 
This detection contains systematic uncertainty due to some dependence on wavelength of the albedo of $\tau$~Bo\"{o}tis~b.  
For the wavelength region of 456--524 nm the planet-to-star flux ratio was found to be $(1.9\pm0.4)\times10^{-4}$, which is brighter than the upper limit set by \citet{Charbonneau1999}. Outside this wavelength region there was no signal detected. 
The wavelength dependence of detected scattered light is useful in investigating the classes of hot Jupiter planet atmospheres. 
Disagreement in planet-to-star flux measurements may be due to the different phase functions adopted for the planets, and the methods used for subtracting the stellar contribution. For example, \citet{Cameron1999} adopt a Venus-like phase function of the planet to account for higher backscattering from clouds, as compared to a simple Lambert-law phase function. 
\citet{Leigh2003a} failed to confirm the tentative detection of light scattered by $\tau$~Bo\"{o}tis b with subsequent reanalysis and additional data.   
\citet{Cameron2000} later revised their result, setting an upper limit on the planet-to-star flux of $\leq 3.5\times10^{-5}$. \citet{Rodler2010} recently measured an upper limit on the light reflected by $\tau$~Bo\"{o}tis b to be $\leq 5.7\times10^{-5}$ times the direct stellar flux, at the 99.9 per cent significance level. This upper limit was based on the method devised by \citet{Charbonneau1999}. 

\citet{Cameron2002} set an upper limit on the $\upsilon$~Andromeda system with a $0.1$~per cent false alarm probability of $\leq5.84\times10^{-5}$.
With the same method, \citet{Leigh2003b} placed an upper limit on the planet-to-star flux of HD~75289b of $\leq4.18\times10^{-5}$ at the $99.9$~per cent level.  However, \citet{Rodler2008} disputed this result, stating that it was based on incorrect orbital phases for the planet. They rederived the upper limit on the planet-to-star flux of HD~75289b to be $\sim60$~per cent higher, via the approach taken by \citet{Charbonneau1999}.  

These methods yield varying results for the same systems, most likely due to the difficulty in measuring the dim planet-to-star flux, 
their model dependence and the unconstrained orbital parameters. However, the upper limits set are lower than predicted by theoretical considerations \citep{Cameron1999}. 
This suggests that hot Jupiter planet atmospheres are less reflective than expected based on the solar system gaseous planets. 
Constraining atmospheric models further requires deeper upper limits on the scattered light,  and minimising the model-dependence of measuring the albedos of hot Jupiter planets. 

In order to reduce the parameter space and number of assumptions, \citet{Liu2007} developed a method specific to detecting the starlight scattered by transiting hot Jupiter planets.  Knowing the planet's velocity from orbital constraints allows a shift-and-add method to be adopted, to verify the presence of the scattered light. Using the equivalent width ratios of the scattered and direct components of the spectra, \citet{Liu2007} found the planet to star ratio of HD~209458b to be $(1.4\pm2.9)\times10^{-4}$ in the wavelength range 544--681 nm.  This result is expected to be updated, following a correction for the non-linearity of the HDS CCDs  \citep{Tajitsu2010}.

Space-based satellites can use transit photometry to measure the starlight scattered by hot Jupiters at a higher  signal-to-noise ratio than possible with ground-based capabilities. 
To determine the amount of starlight scattered by HD~209458b from the depth of the secondary eclipse, \citet{Rowe2008} measured light curves of HD~209458 for the duration of the planetary orbit with the Microvariability and Oscillations of Stars (MOST) satellite.  
Their observations yielded a $1\sigma$ upper limit on the planet-to-star flux ratio of $\leq 1.6\times10^{-5}$.   In the wavelength region of 400--700 nm this corresponds to a geometric albedo of $0.038\pm0.045$ which is much less reflective than the solar system's giant planets. 
This rules out bright clouds at a high altitude, and is consistent with the low albedo upper limits of $\upsilon$ Andromeda and HD~75289 \citep{Cameron2002, Leigh2003b}.  
While searching for transiting planets, the CoRoT and Kepler missions have also yielded a range of planetary upper limits on the albedos of the hot Jupiter planets such as Kepler-7b, Kepler-5b, CoRoT-1b, CoRoT-2b and OgleTr56\citep{Kipping2011, Alonso2009a, Alonso2009b, Rogers2009, Christiansen2010, Sing2009}.
Further studies and repeated observations of these bright, transiting systems will continue to constrain atmospheric models and theories of the formation of solar systems. 
However, space based missions are costly and it is not feasible to consistently follow up on the expected yield of hot Jupiter planets while the search continues for Earth-like extrasolar planets. 
It is therefore beneficial, and timely to develop a suitable model-independent method of extracting the planet-to-star flux from transiting hot Jupiters in ground-based observations. 
It is also advantageous to measure phase independent planet-to-star fluxes without assuming a grey albedo. This motivates a model-independent Fourier based method for measuring the light scattered by hot Jupiter planets, which we introduce in this paper. This method requires minimal assumptions about the system and is suited to transiting planets where parameters are constrained by survey  observations. 

In this paper we quantify the precision of the existing method of \citet{Cameron1999} in measuring the planet-to-star flux of transiting hot Jupiter planets via Monte Carlo simulations. We also introduce and quantify a new method specifically tailored for application to  transiting planets in order to constrain the orbit and detect smaller signals with ground-based capabilities.
We begin with a review of the method of measuring the light scattered by extrasolar planets developed by \citet{Cameron1999}. This method adopts a detailed approach to extracting the signal of the planet from the residual noise of the spectrum by least-squares deconvolution, following the stellar spectrum removal.  Monte Carlo simulations are adopted to determine the sensitivity of this method for application to mock high resolution \'{e}chelle spectra of a transiting hot Jupiter planet system, observed with ground-based capabilities. A new method is then developed which is model-independent and utilises the benefits of reduced parameters for a transiting system. 
This Fourier analysis  method involves solving for the planet and stellar spectra in Fourier space by least-squares, and is applicable to transiting hot Jupiter planets.  To guide future observations, the expected sensitivity to the planet-to-star flux fraction is quantified as a function of   signal-to-noise ratio and wavelength range. This method is then tested on Subaru HDS spectra of HD~209458 and HD~189733, which host a transiting hot Jupiter planet. 

\section{System Parameters}
\label{params}

A planet will reflect an amount ${\textrm f_{{\textrm p}}}$ of its host star's flux ${\textrm f_{\star}}$;
\begin{equation}
 \epsilon(\alpha,\lambda) \equiv \frac{{\textrm f_{{\textrm p}}}(\alpha,\lambda)}{{\textrm f_{\star}}(\lambda)} = {\textrm p}(\lambda){\textrm g}(\alpha,\lambda)\left (\frac{{\textrm R_{{\textrm p}}}(\lambda)}{{\textrm a}}\right )^{2},
\label{epsilon}
\end{equation}
where a is the semi-major axis, ${\textrm R_{{\textrm p}}(\lambda)}$ is the radius of the planet, ${\textrm p}(\lambda)$ is the geometric albedo, and ${\textrm g}(\alpha, \lambda)$ is the phase function at  planetary phase angle $\alpha$ and wavelength $\lambda$ \citep{Leigh2003c}.  
From theoretical considerations a planet-to-star flux of $\epsilon \sim 10^{-4}$ is expected for a hot Jupiter planet at full phase \citep{Cameron1999}.  

\subsection{Phase Function}
\label{phase}

The flux scattered by a hot Jupiter planet is a phase dependent fraction of the maximum reflectance at full phase.  
Unfortunately, due to a lack of signal and phase coverage it is not yet possible to determine the exact phase function ${\textrm g}(\alpha)$ for a hot Jupiter planet being observed, and if required, it must therefore be modelled as a function of the phase angle $\alpha$. 

It is simplest to adopt a Lambert-law sphere which assumes isotropic scattering over $2\pi$ steradians. This typically has the form;
\begin{equation}
 {\textrm g}(\alpha) = [\sin \alpha + (\pi - \alpha)\cos \alpha] / \pi,
\end{equation}
where the phase angle $\alpha$ is determined from the inclination $i$ and the orbital phase $\phi$;
\begin{equation}
\cos \alpha = {\textrm -} \sin i \cos 2 \pi \phi.
\end{equation}
For an orbital phase of $\phi  = 0$ the planet is closest to the observer, and has the smallest planetary phase. For an orbital phase of $\phi  = 0.5$, the planet is at the furthest point in its orbit from the observer, with maximum reflectance.

To account for stronger back-scattering than the Lambert-law phase function, it may be more realistic to adopt the phase function of a similar planet in the Solar System, that has a cloud-covered surface. 
For example; 
\begin{equation}
  {\textrm g}(\alpha) = 10^{-0.4\Delta {\textrm m}(\alpha)},
  \label{eqn_planet_phase}
\end{equation}
where;
\begin{equation}
  \Delta {\textrm m}(\alpha) = 0.09\left(\frac{\alpha}{100^{o}}\right) + 2.39\left(\frac{\alpha}{100^{o}}\right)^{2} - 0.65\left(\frac{\alpha}{100^{o}}\right)^{3},
\end{equation}
which is a polynomial approximation to the empirically determined Venus-like phase function \citep{Cameron2002}.
For transiting planets, which have orbital inclinations close to $90^{o}$, these two functions have a similar variation in flux of less than 10~per cent  over the orbital phases adopted for the mock spectra presented here, and the Subaru~HDS spectra of HD~189733b and HD~209458b.

\subsection{Velocity}

A planet's velocity relative to its host star  ${\textrm V_{{\textrm p}}}(\phi)$ at an orbital phase $\phi$ is given by;
\begin{equation}
{\textrm V_{{\textrm p}}}(\phi) = {\textrm K_{{\textrm p}}}\sin(2\pi\phi),
\label{vel}
\end{equation}
where $\phi = ( t - {\textrm T_{0}})/{\textrm P_{{\textrm o \textrm r \textrm b}}}$ at the time $t$, where ${\textrm T_{0}}$ is the transit epoch and ${\textrm P_{{\textrm o \textrm  r \textrm  b}}}$ is the orbital period, for a circular orbit (ellipticity e=0). The apparent radial-velocity amplitude ${\textrm K_{{\textrm p}}}$ about the centre of mass of the system is;
\begin{equation}
{\textrm K_{{\textrm p}}} = \frac{2\pi {\textrm a}}{{\textrm P_{{\textrm o\textrm r\textrm b}}}}\frac{\sin i}{1+{\textrm q}},
\end{equation}
where a is the orbital distance, $i$ is the inclination and ${\textrm q} = {\textrm M_{{\textrm p}}}/{\textrm M_{\star}}$ is the mass ratio of the planet to the star.   For a system under consideration, a, $i$ and ${\textrm R_{{\textrm p}}}$ can be determined from radial-velocity measurements and transit photometry. For non--transiting planets these parameters have greater uncertainty, and ${\textrm K_{{\textrm p}}}$ is determined from fitting observational light curves.  

\section{Construction of Mock Spectra}
\label{testspectra}

\begin{figure*}
\centering
\begin{minipage}{160mm}

\begin{center}
  \includegraphics[width=120mm]{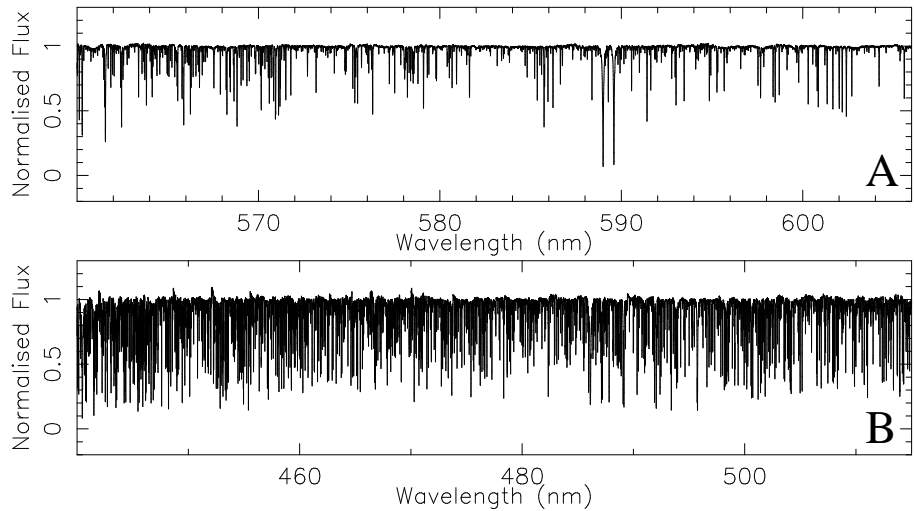}
 \caption[Example mock spectra]{{\bf Mock spectra constructed from high
 signal-to-noise ratio solar spectrum.} Panel {\bf A} - Example mock
 spectra for red wavelength range of 560--608 nm. Panel {\bf B} - Example mock spectra for blue wavelength range of  440--515 nm. Statistical Gaussian noise added to the spectrum is distributed with standard deviation $\sqrt {\textrm N}$ for N counts per pixel. The wavelength is logarithmically binned such that the Doppler shifted scattered light can be added as an integer pixel shifted copy of the star, with amplitude scaled by $\epsilon  {\textrm g}(\alpha)$. The orders of the spectra were combined and normalised with an 11-knot cubic-spline.\label{testsp}}
\end{center}
\end{minipage}
\end{figure*}

Mock spectra were constructed in order to quantify the precision of methods for measuring the light scattered by extrasolar planets, including that by \citet{Cameron1999}, as well as the new Fourier analysis method, introduced in Section 5. 
Spectra were constructed from high resolution solar spectra ($\lambda/\Delta\lambda~\sim~300,000$), obtained with the Fourier Transform Spectrometer at the McMath/Pierce Solar Telescope situated on Kitt Peak, Arizona\footnote{Operated by the National Solar Observatory, a  Division of the National Optical Astronomy Observatories.  NOAO is  administered by the Association of Universities for Research in Astronomy, Inc., under cooperative agreement with the National Science Foundation.}.
The mock spectra were constructed to be analogous to observing a bright hot Jupiter transiting system, such as HD~209458, with typical  ground-based 8~m telescope capabilities, such as Subaru~HDS. An example mock spectrum is shown in Fig. \ref{testsp}.
The high resolution solar spectra were scaled to the counts expected from HD~209458 with Subaru~HDS for a typical exposure time of 500 seconds. A spectral resolution of R~$=\lambda/\Delta\lambda=45,000$ was mimicked by Gaussian smoothing, consistent with twice the critical oversampling. Noise due to statistical fluctuation in photons from the hypothetical measurement with Subaru~HDS was added after including the planet contribution and smoothing.  The noise is Gaussian distributed with a standard deviation of $\sqrt {\textrm N}$ for N counts per pixel. 
The planet velocity dictates the wavelength shift of the scattered spectrum, taken as a scaled copy of the stationary stellar spectrum. During the length of an exposure the planet velocity varies by less than the width of the absorption features. 

Planet velocities were randomly selected within the range 30--80~${\textrm  k \textrm m \ \textrm s^{-1}}$ towards the observer. This is consistent with the range of velocities of HD~209458b when the planet is just out of secondary eclipse and the orbital phase is in the range 0.55--0.60.  There is less than 10 per cent variation in the flux scattered from the planet within this phase range based on either a Lambert-sphere or a Venus-like phase function. 
This is a benefit of using transiting systems to measure planet-to-star fluxes. As the planet velocity amplitude can be measured from the constrained parameters, a smaller range of the orbit can be observed and selected to be where the planet is close to full phase and maximum reflectance. The velocities must be large enough to shift the planet signal clear of the stellar absorption lines. 

Instrumental variations, such as flexure of the spectrograph and guiding errors, can cause wavelength variations between the spectra during a night of observations \citep{Winn2004}. This can be accounted for in data reduction by adjusting the wavelength calibration using the star's spectral absorption lines. Variation in the flux calibration between exposures is removed by continuum fitting and normalisation. 
The mock spectra are taken to be ideal, with no variation in the shape of the response of the CCD with time and the host star spectra in all  exposures are assumed to be perfectly aligned.

\section{Collier Cameron Method}
\label{colliercameron}

The most widely used current method (henceforth the Collier Cameron method) was first presented by \citet{Cameron1999}, and further developed by \citet{Cameron2002}  and \citet{Leigh2003b}.  
The Collier Cameron method has previously been adopted to attempt to measure the light scattered by three different, non-transiting hot Jupiter planets \citep{Cameron1999, Cameron2002, Leigh2003b}. It involves detailed modelling of the stellar contribution and matched-filter analysis to extract the planet signal from the noise. It is a useful method for the case where the hot Jupiter planet's orbit is unknown, and has allowed constraints to be placed on highly reflective planetary atmospheres \citep{Rodler2010}.
In this Section, the Collier Cameron method is reviewed for application
to a transiting planet system, using the mock spectra described in
Section~\ref{testspectra}, within Monte Carlo simulations.  

For the analysis of non-transiting planets, the Collier Cameron method
requires observations of as much of a full orbit as possible, such that
the most likely orbital phase function and planet velocity can be
measured.  For transiting systems the velocity of the planet is known,
and a small section of the orbit can be observed. As we are evaluating
the Collier Cameron method for the case where the planet is transiting,
the thirty mock spectra that were used to measure a planet-to-star flux
are in a small orbital phase range, either side of the secondary eclipse,
 when  the majority of the dayside of the planet is facing the observer. 
This corresponds to velocities with magnitudes of 30--80~${\textrm  k \textrm m \ \textrm s^{-1}}$.
The planet's spectral features are therefore shifted clear of the corresponding stellar features, and the planet is close to maximum brightness.
The planet is assumed to follow a Venus-like phase function as defined in Section~\ref{phase}. 
The initial input spectra do not need to be merged for this method, but
can remain as separate orders, without continuum normalisation. The
spectra are in the red  wavelength range of 560--680 nm, and the blue wavelength of range of 440--515 nm.
We note that the mock spectra are ideal, in that they are all aligned, without cosmic ray hits or vignetting, and the stellar contribution is constant.

\begin{figure}
\begin{center}
  \includegraphics[width=80mm]{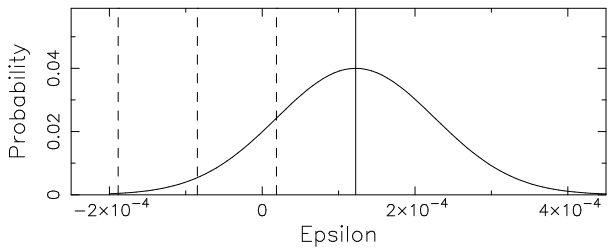}
 \caption[Probability distribution]{{\bf Example probability
 distribution.} An example probability distribution with a mean planet-to-star flux of $\epsilon= 1.2\times10^{-4}$ marked by the vertical solid line. The distance of one, two and three standard deviations from the mean are marked by the vertical dashed lines. This distribution has a statistical significance of $>1\sigma$ as the mean is greater than one standard deviation, but less than two standard deviations from zero. \label{sigma}}
\end{center}
\end{figure}

\begin{figure*}
\centering
\begin{minipage}{160mm}
\begin{center} 
  \includegraphics[width=150mm]{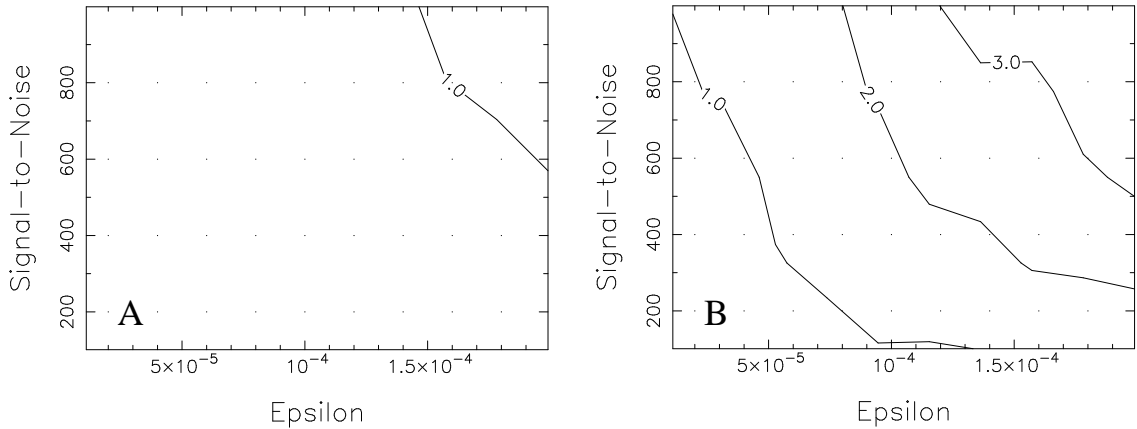} 
 \caption[Monte Carlo results for red and blue wavelength ranges]{{\bf
 Signal-to-noise ratio for mock spectra with Collier Cameron method.}
 Contour plot of $1\sigma$,
 $2\sigma$ and $3\sigma$ levels of detection  (solid)  for
 signal-to-noise ratio values in the range $10^{2}$--$10^{3}$, and
 $\epsilon$ values in the
 range $10^{-5}$--$2\times10^{-4}$.  
 Panel {\bf A} - Results for the red wavelength range
 560--680~nm. Panel {\bf B} - Results for the blue wavelength range
 440--515~nm. The dots represent the sampling of the parameter space.  \label{ccplot}}
\end{center}
\end{minipage}
\end{figure*}

The Collier Cameron method requires modelling and removing the direct starlight from the total spectrum, leaving the planet signal buried in noise. It is very difficult to accurately remove the stellar contribution, as small distortions in the strong stellar lines can produce changes larger than the planet signature. Problems can arise due to shifting of the spectrum on the detector, changes in the telescope focus and atmospheric seeing. In the blue wavelength range, the density of strong lines makes removing the stellar signal particularly difficult, and observations for previous upper limit measurements have generally been in a central wavelength range. 

In this paper, results are presented for the ideal mock spectra, and the stellar spectrum does not vary with exposure. This aids the removal of the modelled stellar spectrum, since we do not have to account for sources of distortion. This simplification of the removal of the direct light is equivalent to perfectly modelling the stellar spectrum, and is the best case scenario for the Collier Cameron method described in \citet{Cameron2002}.  The sensitivity of the method to the planet-to-star flux will be impeded by instrumental variations, and the difficulty in modelling the stellar spectrum, particularly in the blue wavelength. 

The planet signal strength is extracted from the residual after the direct flux subtraction via least-squares deconvolution with a set of known weights of stellar absorption lines. The method is described in detail in \citet{Cameron2002}. This results in a deconvolved velocity profile which represents the lines present in the planet spectrum. It is similar to a cross-correlation, but neighbouring blended lines are dealt with more accurately. For the mock spectra, a library of spectral lines in the required wavelength range was compiled from data for the Sun from the Vienna Atomic Line Data Base (VALD)  \citep{Kupka1999}. 

There are typically ripples in the deconvolved stellar profile at low velocities that may cause spurious effects on measuring the planet contribution \citep{Cameron2002}.  For the Monte Carlo simulations the planet signal was extracted from the range of velocities with magnitudes greater than $30~{\textrm  k \textrm m \ \textrm s^{-1}}$. To extract the planet signal, a matched-filter is constructed, and fit to the data to determine the planet-to-star flux. The velocity profile of the planet signal is modelled as a moving Gaussian, based on the form used in \citet{Cameron2002}. The planet-to-star flux is determined by scaling the matched filter to match the deconvolved planet line profile, and finding the best fit with a $\chi^{2}$ minimisation \citep{Cameron1999}.

\subsection{Monte Carlo Analysis}

The sensitivity of the Collier Cameron method was determined for a range of  signal-to-noise ratio strengths and planet-to-star fluxes via Monte Carlo simulations. 
The typical  signal-to-noise ratio per pixel of the mock spectra was varied by increasing the counts per pixel, corresponding to a longer exposure time.  A larger telescope collecting area or observing a brighter star would also increase the counts per pixel.
The number of standard deviations separating a detection from zero was determined by the mean of the distribution of all Monte Carlo simulation solutions, divided by the standard deviation.
This defines the distance of the mean value from zero in units of the spread of the distribution.  For example, a result quoted at $1\sigma$ has a distribution with a mean value that is one standard deviation away from zero, as shown in Fig. \ref{sigma}.

In the red wavelength range of the mock spectra of 560--680~nm, with
1465
absorption lines, a signal-to-noise ratio of $\sim 800$ is required for
a $1\sigma$ detection of a planet-to-star flux of $1.6\times10^{-4}$.  
The blue wavelength range of 440--515~nm has many more absorption lines
(2570), such that the method is more sensitive to smaller planet-to-star
fluxes for data with the same signal-to-noise ratio, assuming the
stellar spectra is perfectly removed.  
Planet-to-star fluxes greater than $10^{-4}$ are detectable at the
$1\sigma$ level. 
A planet-to-star flux of $1.6\times10^{-5}$
requires a signal-to-noise of around $\sim 800$ to be detected at the
$3\sigma$ level, for the ideal case. 
The results of the Monte Carlo simulations with varying signal-to-noise
ratio for the red wavelength range of 560--680~nm and blue wavelength
range of 440--515~nm are presented in Fig.~\ref{ccplot}. 
There is a limit on the possible   signal-to-noise ratio of the spectra due to the capability of the CCD and variations in the planet velocity during longer exposure times smearing the scattered signal. 

The results of the Monte Carlo simulations for ideal spectra are consistent with
detections via this method for real spectra. Probable detections
have been made for planet-to-star fluxes brighter than  $\epsilon = 10^{-4}$, and upper limits are set with low false-alarm probabilities.
For example, \citet{Cameron1999} detected a best-fit to the signal of
the light scattered by $\tau$ Bootis b of $\epsilon = 1.9\times10^{-4}$
for data with a typical   signal-to-noise ratio of $\sim$1000, in a central wavelength range. 

As the Collier Cameron method was developed prior to the discovery of
transiting hot Jupiter planets, it did not take advantage of the
minimisation of unknown parameters in these systems. This method can
benefit from the known planet velocity in extracting the scattered
signal.  However in constructing the template spectrum, the spectra are summed to smear out the planet signal. This requires a large range of velocities, or isolating spectra with the planet at minimal illumination. Observing a small section of the transiting planet orbit may therefore affect the accuracy of the stellar model, and reduce the planet signal in the residual spectrum. 

While the analysis methods developed by \citet{Charbonneau1999} and \citet{Cameron1999} are required for a non-transiting system, 
they have not yielded a detection of the light scattered by a hot Jupiter planet.  With the focus of many current extrasolar planet surveys on transiting objects, it is therefore timely to develop a method of extracting the planet-to-star flux specific to transiting hot Jupiter planets, that does not require modelling the stellar contribution. In the next Section we describe a model-independent Fourier analysis method that requires minimal assumptions about the system and is suited to transiting planets where the parameters are constrained by other space-based, survey observations.

\section{Fourier Analysis Method}
\label{jenkins}

\begin{table*}
\centering
\begin{minipage}{120mm}
\begin{tabular}{cccccc}
\hline
& \multicolumn{4}{c}{${\textrm C_{i,j}}$}& \\
\cline{2-5}
  & j=1 & j=2 & j=3 & j=4 & $b_{i}$ \\
\hline
&&&&& \\
 i=1 & n & 0 & $\sum\cos\Delta_{k}$ & $-\sum\sin\Delta_{k}$ &  $\sum t_{x,k}\cos\Delta_{k} - \sum t_{y,k}\sin\Delta_{k}$   \\
&&&&& \\
i=2 & 0 & n & $\sum\sin\Delta_{k}$ & $\sum\cos\Delta_{k}$ &  $\sum t_{x,k}\sin\Delta_{k} + \sum t_{y,k}\cos\Delta_{k}$   \\
&&&&& \\
 i=3 & $\sum\cos\Delta_{k}$ & $\sum\sin\Delta_{k}$ & n & 0 &  $\sum t_{x,k}$ \\
&&&&& \\
 i=4 & $-\sum\sin\Delta_{k}$ & $\sum\cos\Delta_{k}$ & 0 & n &  $\sum t_{y,k}$ \\
&&&&& \\
\hline
\end{tabular}
 \caption[Coefficients for the matrix {\bf C}($\omega$) and vector {\bf b}($\omega$)]{{\bf Coefficients for the matrix {\bf C}($\omega$) and vector {\bf b}($\omega$)}. All summations are from k = 1 to n, where n is the number of spectra.\label{cmatrix}}
\end{minipage}
\end{table*}

The Fourier analysis method for extracting the scattered signal from the combined planet and star flux described in this Section is based on a method developed to deal with the effect of the fixed-pattern response function of a detector on a moving spectrum \citep{Jenkins2002}.  In the case of a transiting planet, the stellar spectrum is considered to be stationary (analogous to the response function), while the planet's scattered spectrum moves with each exposure.  
Separating out a moving spectrum via a Fourier transform is a useful technique for 
binary systems (see e.g. \citet{Hadrava1995}). 
The observed spectrum recorded by the \'{e}chelle spectrometer ${\textrm T}(\lambda)$, consists of the direct starlight contribution ${\textrm S}(\lambda)$, plus the starlight scattered by the orbiting planet ${\textrm P}(\lambda)$.  The planet spectrum is shifted with respect to the stationary stellar spectrum with a predictable Doppler shift, dependent on the velocity of the planet at time of the exposure.  There is also a contribution from statistical noise that is not frequency dependent. 

For the total spectrum ${\textrm T}(\lambda)$ we define:
\begin{eqnarray}
	 t_{x}(\omega) = Re\{\mathscr{F}[{\textrm T}(\lambda)]\}, \\ 
         t_{y}(\omega) = Im\{\mathscr{F}[{\textrm T}(\lambda)]\},
\label{ft}
\end{eqnarray}
where $\mathscr{F}$ is the Fourier transform operator.  Similarly
$p_{x}(\omega)$, $p_{y}(\omega)$, $s_{x}(\omega)$ and $s_{y}(\omega)$
are defined for the planet and stellar spectra. As each spectrum has a
different Doppler velocity displacement for ${\textrm P}(\lambda)$, depending on the observed velocity of the planet, the zero shift complex function $p_{x}(\omega) + i p_{y}(\omega)$ is multiplied by $\exp(- 2 \pi i \omega \delta_{k})$, where $\delta_{k}$ is the magnitude of the shift of ${\textrm P}(\lambda)$ in the $k^{th}$ exposure compared to the zero velocity case.

The minimum of the sum of the squared real and imaginary residuals (the minimum of ${\textrm Q^{2}}$) is solved to determine the unknown planet and stellar spectrum coefficients $p_{x}(\omega)$ and $p_{y}(\omega)$, $s_{x}(\omega)$ and $s_{y}(\omega)$.  This is done by setting the partial differentials of ${\textrm Q^{2}}$ defined below with respect to the four unknowns  $p_{x}$, $p_{y}$, $s_{x}$ and $s_{y}$ at each $\omega$, to be zero;
\begin{eqnarray}
{\textrm Q^{2}}  =  \sum_{k=1}^{n}(p_{x}\cos\Delta_{k} + p_{y}\sin\Delta_{k} + s_{x} - t_{x,k})^{2} \\
+ \sum_{k=1}^{n}(-p_{x}\sin\Delta_{k} + p_{y}\cos\Delta_{k} + s_{y} - t_{y,k})^{2} ,
\label{sumsq}
\end{eqnarray}
where $\Delta_{k}$ is an abbreviation for $2 \pi \omega \delta_{k}$.

Therefore, the best solutions for $p_{x}(\omega)$, $p_{y}(\omega)$, $s_{x}(\omega)$ and $s_{y}(\omega)$ are found by solving a system of four linear equations at each $\omega$. The linear equations can be summarised as;
\begin{equation}
	{\bf C}(\omega)\cdot{\bf u}(\omega) = {\bf  b}(\omega) ,
\label{lineareq}
\end{equation}
where {\bf u}($\omega$) is a vector of $p_{x}$, $p_{y}$, $s_{x}$ and $s_{y}$ at each $\omega$ and the coefficients ${\textrm C_{i,j}}$ and ${\textrm b_{i}}$ are listed in Table~\ref{cmatrix}.

 Singlular value decomposition is used to solve the system of linear equations. It can be verified that the solution is a minimum in ${\textrm Q^{2}}$ by plotting nearby solutions.  After solving for each vector {\bf u}($\omega$) over all $\omega$, the inverse Fourier transforms of the two pairs of terms $u_{1}(\omega) + iu_{2}(\omega)$ and $u_{3}(\omega) + iu_{4}(\omega)$ recover the best representations of ${\textrm P}(\lambda)$ and ${\textrm S}(\lambda)$.  The planet spectrum, ${\textrm P}(\lambda)$, is divided by the stellar spectrum, ${\textrm S}(\lambda)$, and the mean found over all wavelengths, yielding the corresponding planet-to-star flux.
The solution can be verified to be the planet contribution by repeating the measurement while shuffling the order of the Doppler shifts with respect to their exposures, which will yield a null result.

If the velocity displacements of the
observations in a set are equally spaced, 
the the matrices become singular and the solutions can blow
up. Therefore, the timing of the observations should be such that the
  velocity shifts are noncommensurate. This is not unique to the Fourier
  analysis method, and should be considered when planning an observing program.

Solving for the planet and stellar spectra in this way has the following benefits; (i) it does not require a model of the star, 
(ii) it is still effectively using the signal of all of the absorption lines, although they are not summed over to increase the   signal-to-noise ratio, meaning that the planet-to-star flux solution can be wavelength dependent, and
(iii) there is no assumption about the form of the planet spectra, only that it is moving compared to the stationary stellar spectrum with specific Doppler shifts. For example, forcing the planet spectra, $p_{x}(\omega)$ and $p_{y}(\omega)$ to be a scaled copy of the stellar spectra, $\epsilon(\alpha) s_{x}(\omega)$ and $\epsilon(\alpha) s_{y}(\omega)$ respectively, would require solving three highly non-linear coupled equations, and remove the freedom of solving for a wavelength dependent albedo. 

\subsection{Fourier Transform}
\label{fourier}

\begin{figure}
\begin{center}
  \includegraphics[width=80mm]{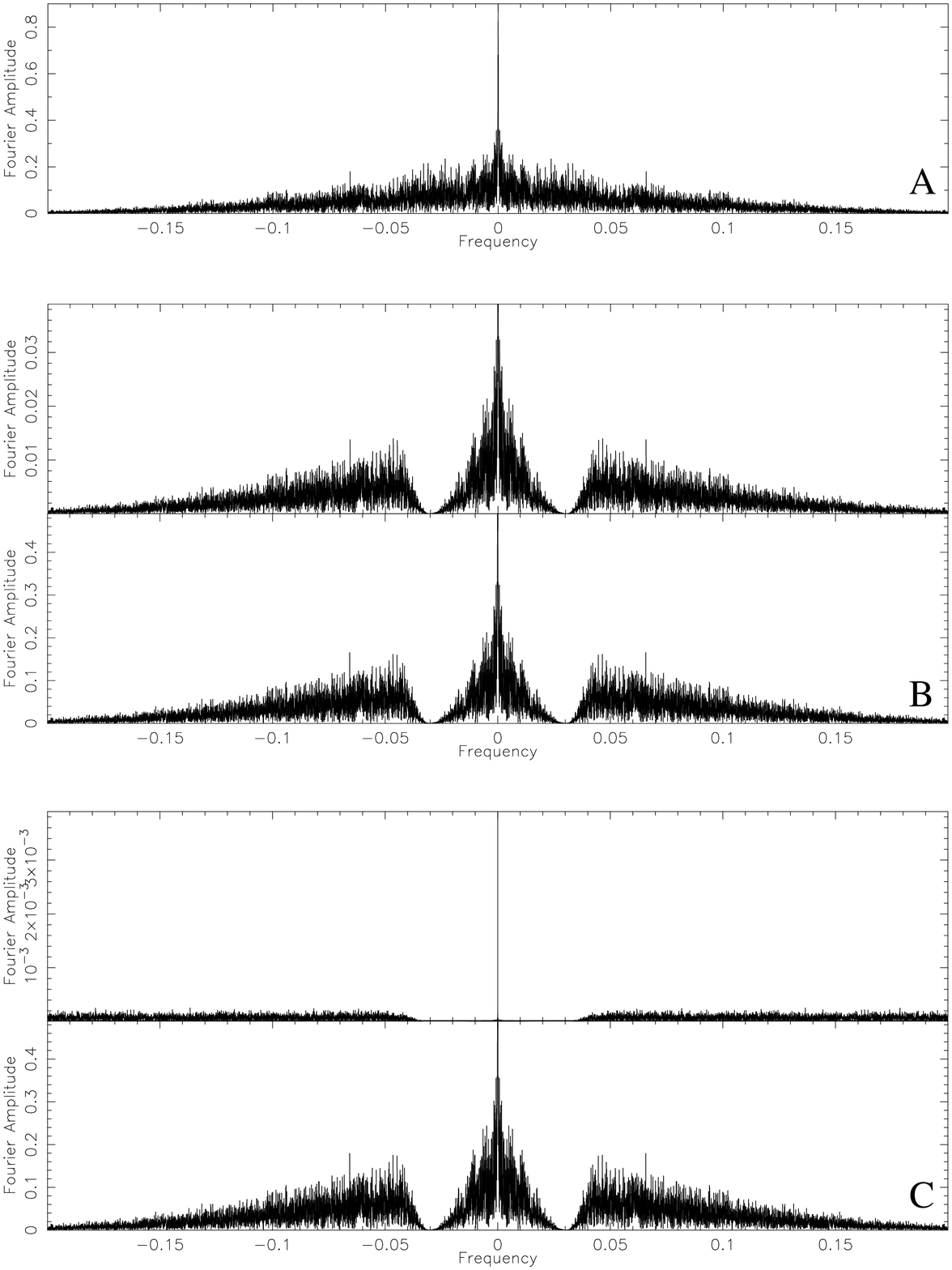}
 \caption[Damping spurious signals at low frequencies.]{{\bf Damping
 spurious signals at low frequencies.} Panel {\bf A} - Example of
 Fourier amplitudes of input blue mock spectra. Panel {\bf B} - Example
 of Fourier amplitudes of the solution blue planet ({\it upper
 subpanel}) and star ({\it lower subpanel}) for mock spectra with
 $\epsilon$ = 0.1 and low frequencies replaced with the artificial
 signal to damp spurious small frequency signals. For the case where the
 planet-to-star flux is large, the planet and stellar signals appear to
 be the same shape. Panel {\bf C} - As for Panel B, with $\epsilon$ =
 0.0001. For the smaller $\epsilon$ value the moving signal solution is
 of the order of the statistical noise, and hence the planet signal
 appears to be flat out to large frequencies. The central, small
 amplitude regions in Panels B and C, are due to the need for a smooth
 transition between the solution signal and the artificial signal. In
 this case a Hanning window was used to damp the solution and artificial
 signals to zero where they meet. The frequency range lost in doing this
 is the same for the planet and the star, therefore the resulting
 planet-to-star flux will not be affected. This can be verified with the
 mock spectra generated with particular $\epsilon$ values. The Fourier
 amplitude of the zero frequency extends above the scale plotted, but
 has a finite value. Note that only every 5th point is plotted here. \label{damping}}
\end{center} 
\end{figure}

The Fourier transform is computed using FFTW, a C subroutine library \citep{FFTW}\footnote{See htp://www.fftw.org/}. Before transforming the input spectra into Fourier space, the spectra are extended above and below the wavelength range by adding the value of the continuum, and then damped to zero at the ends gradually, using a Hanning window which has the form;
\begin{equation}
f(\lambda) = \frac{1}{2}\left(1 - \cos\left(\frac{2\pi \lambda}{{\textrm N}}\right)\right),
\end{equation}
for N pixels.
FFTW assumes that the spectrum is infinitely periodic, and the Hanning window suppresses discontinuities as it loops over the wavelength range when transforming into Fourier space. 

The low frequency results of $p_{x}(\omega)$ and $p_{y}(\omega)$,
$s_{x}(\omega)$ and $s_{y}(\omega)$ need to be dealt with carefully. 
At these frequencies the phase shifts induced by the changes in the planet's velocity become so 
small that they are difficult to detect. In essence, the planet and star signals are no longer
easily separable.  Mathematically, this problem manifests
itself by making the matrix C singular or nearly so.
As a consequence, the solutions start to be dominated by
noise or low-frequency systematic errors.  By contrast,
narrow stellar features that have a presence in both
${\textrm S}(\lambda)$ and ${\textrm P}(\lambda)$ can create measurable amplitudes
and phase shifts at high frequencies, and thus they contain
most of the planet's detectable signal.
In order to extract this important information, the spectra are effectively put through a high-pass filter, and the low frequencies are not solved for. 
So that the result can be inverse Fourier transformed, the low frequency
solutions in Fourier space are replaced with a predicted solution. This
requires the simple assumption that the planet spectrum is a scaled copy
of the stellar spectrum. 
The average of the Fourier amplitudes of the input spectra is taken to be an estimate of the Fourier amplitudes of the stellar spectra. The planet signal will be averaged out, due to different Doppler shifts in each exposure. This is multiplied by the predicted amplitude of the planet-to-star flux, and this test amplitude is iterated until it equals the solution from the Fourier analysis method - the real planet-to-star flux. A null result occurs when this iteration does not converge.

The frequency region of the solution which is artificially replaced is determined for when the matrix {\bf C} is singular ($\omega = 0$) or nearly singular.  The gradual transition at the edge of this region uses a Hanning window to damp the artificial and solution Fourier amplitudes to zero where they meet (see Fig.~\ref{damping}). This is not the only possible method to combine the artificial and real solutions, however it must be gradual  enough not to cause ringing in  the planet and star spectra when they are transformed back to real space.  Due to this transition, sections of frequency space are lost. However, as it occurs equally in the stellar and planet spectra (due to the matching transition range), the resulting planet-to-star flux value is conserved. 
If the Fourier components are used to compare the stellar and planet solutions, this artificial replacement is not required, however the low frequency result is lost. 

The central frequency cannot just be set to zero to remove the low frequency solutions if the inverse Fourier transform is required, as it fixes the overall continuum average to also be zero.

\subsection{Measuring the Planet-to-Star Flux}

The planet and the stellar spectra are solved for at each wavelength, by taking the inverse Fourier transform of the resulting Fourier components $p_{x}(\omega)$ and $p_{y}(\omega)$, $s_{x}(\omega)$ and $s_{y}(\omega)$.
The spectra are retrieved, including the Hanning-windowed continuum-added ranges either side of the pixel range corresponding to the solution planet and stellar spectra; ${\textrm P}(\lambda)$ and ${\textrm S}(\lambda)$.
The planet spectrum solution is aligned with the stellar spectrum, as it is shifted to the zero velocity case.

The zero frequency value of the spectra is the power in the longest wavelength Fourier component, F(0). This corresponds to the continuum level, however we cannot explicitly solve for the zero shift solution. Therefore, it is artificially constructed before the inverse Fourier transform, and is used to iterate to the correct solution. 
The planet spectrum is divided by the stellar spectrum, and the mean ratio value taken for the entire range to be the value of the planet-to-star flux. This is then tested against the test amplitude for the artificial low frequency solutions, until the two values converge on the solution.

The planet-to-star flux of the observation is determined with a small range of orbital phases, so that the amount of light the planet reflects does not vary rapidly. In order to scale this to the maximum reflectance for comparison with the Collier Cameron result, a planet phase model must be assumed.  For the mock spectra, the orbital phase is in the range 0.55--0.60. This corresponds to an average planet phase of 85~per cent illumination, for a Venus-like phase function. The planet phase at each exposure cannot be accounted for in solving for the planet spectrum in Fourier space, requiring a small region of the orbit to be observed. The actual planet-to-star flux $\epsilon(\alpha)$ is therefore the model-independent measurement, which will be less then the maximum reflectance upper limit.

\subsection{Non-Grey Albedo}

So far in this paper, a grey albedo has been assumed in constructing
mock spectra. This means that the planet reflects an equal fraction of
the starlight across the entire wavelength range. In the Collier Cameron
method, the line weights can be attenuated according to a particular
atmosphere model to account for a non-grey albedo. For exanple, in order to test for different classes of planets, such as the Class V
roaster or Isolated Class IV models, a wavelength dependent
planet-to-star flux cold be used for the artifical solution. Any increase in the likelihood of the measured planet-to-star flux can be used as an indication of the accuracy of the model. 

The Fourier analysis method provides an opportunity to solve for the planet signal at each wavelength, and therefore determine any slope of the albedo function with wavelength. However, as the zero shift solution cannot be solved, the F(0) values are estimated based on the assumption that the stellar spectrum can be represented by the average input spectrum, and the planet as a scaled copy. Therefore, varying the shape of the continuum of the resulting planet spectrum requires modelling possible wavelength dependent albedos as an iterative parameter to find the best solution. It is also possible to divide the input spectrum into short wavelength regions over which a non-grey albedo would vary slowly. This was the process adopted by \citet{Cameron1999} for $\tau$~Bo\"{o}tis~b over ranges of 40~nm, and their results suggested a wavelength dependence of the albedo at around 500~nm.
Alternatively, solving for the planet's scattered spectrum allows the chance for strong absorption lines to be explicitly revealed in high signal-to-noise ratio data. Typically these lines are expected to be below the level required to detect single absorption lines, but an advantage of  the Fourier method is that it  preserves the details of the planet's spectrum.

\subsection{Monte Carlo Analysis}
\label{mc}

 Monte Carlo simulations were run for the Fourier analysis method with the mock spectra in order to determine the observing strategy most sensitive to small planet-to-star fluxes, and improvements possible with minimal assumptions and typical transit data. 

\subsubsection{ Signal-to-Noise Ratio}
\label{SNR}

\begin{figure*}
\centering
\begin{minipage}{160mm}

\begin{center}
  \includegraphics[width=150mm]{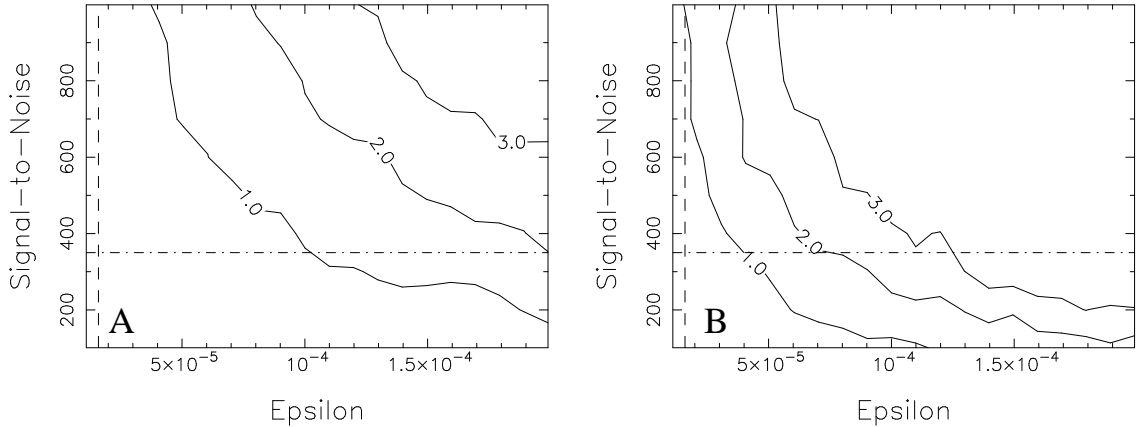}
 \caption[  signal-to-noise ratio for mock spectra in red CCD]{{\bf
 Signal-to-noise ratio for mock spectra with the Fourier analysis
 method.} Contour plot of
 1$\sigma$, 2$\sigma$ and 3$\sigma$ levels of detection (solid) for
 signal-to-noise ratio values in the range $10^{2}$--$10^{3}$, and
 $\epsilon$ values in the
 range $10^{-5}$--$2\times10^{-4}$. {\bf Panel A} - Results for the red wavelength range
 560--608 nm. {\bf Panel B} - Results for the blue wavelength range 440--515 nm. 
 The dots represent the sampling of the parameter space. 
 The dot-dashed line
 shows the signal-to-noise ratio value of 350 for the HD 209458 data.
 A $1\sigma$ level of detection for this data requires a planet-to-star
 flux greater than $\epsilon = 1\times10^{-4}$ in the red wavelength
 range,
 and $\epsilon = 4\times10^{-5}$ in the blue wavelength range. The dashed line
 shows the upper limit on the planet-to-star flux of HD~209458b set by
 \citet{Rowe2008} of $\epsilon<1.6\times10^{-5}$. \label{Fourierplots}}
\end{center}
\end{minipage}
\end{figure*}

We constructed spectra with a range of signal-to-noise ratio and
planet-to-star flux ratios, as per the Monte-Carlo analysis discussed in
Section 4.1.
The results of the Monte Carlo simulations with varying
signal-to-noise ratios are presented in Fig.~\ref{Fourierplots},
corresponding to the red wavelength region of 560--608 nm and the blue
wavelength region of 440--515 nm respectively. The planet-to-star flux
is theoretically predicted to be $\epsilon \leq 10^{-4}$
\citep{Cameron1999}; the range of $\epsilon$ values is centered on this value.
Fig.~\ref{sigma} shows the probability distribution and sensitivity
level determined from an example Monte Carlo simulation, centred on the expected planet-to-star flux at maximum reflectance for the system.
The resulting  planet-to-star fluxes are scaled to the maximum
reflectance for the contour plots presented in Fig.~\ref{Fourierplots}.

The method is more sensitive to smaller planet-to-star fluxes in the
blue wavelength range than the red wavelength range. This is most likely
due to the increased number of absorption lines that have a larger
signal in the Fourier domain, and in the higher frequency ranges, not as
affected by small shifts in the planet spectra.
For a typical   signal-to-noise ratio of around 350, a 1$\sigma$
detection can be made of scattered light  with  $\epsilon \geq
4\times10^{-5}$ in the blue wavelength region, as compared with
$\epsilon \geq 1 \times 10^{-4}$ for the red wavelength region of the
spectrum.  In Fig.~\ref{Fourierplots} the contours of levels of
detection have a very steep gradient as $\epsilon$ decreases to
$10^{-5}$, suggesting that very dim planets will require very high
signal-to-noise ratios to be detectable via this method and ground-based
capabilities. 
The theoretically predicted planet-to-star flux of $\epsilon \leq 10^{-4}$
\citep{Cameron1999} is detectable at the $3\sigma$ level with reasonable signal-to-noise values
in the blue wavelength region, and at the $1\sigma$ level in the red
wavelength region. 
Similar to the Collier Cameron result, 
matching the $1\sigma$ upper limit of HD~209458b measured with the
space-based satellite MOST \citep{Rowe2008} requires the blue wavelength
range Subaru~HDS spectra with a  signal-to-noise ratio greater than
$\sim900$. 

\subsubsection{Spectral Resolution}

The spectral resolution of the test input spectra was also varied for Monte Carlo simulations with the Fourier analysis method. The   signal-to-noise ratio was kept constant per pixel, but the oversampling varied by degrading the spectrum with Gaussian smoothing over an increasing radius.
The spectral resolution range of R = 32,000--90,000 corresponds to slit widths of 1.2--0.4'' and an oversampling of 4--12 pixels per resolution element. The critical sampling is 2 pixels per resolution element for Subaru~HDS. 

Within this range there is no improvement in the sensitivity of the
Fourier analysis method.  However, increased spectral resolution may be
beneficial to the data reduction process and correcting for instrumental
effects and wavelength shifts not incorporated into the ideal test
spectra. On the other hand, the increase in signal-to-noise ratio in
\'{e}chelle data gained by a decrease in spectral resolution may yield a
measurement of a smaller planet-to-star flux. 

\subsubsection{Wavelength Range and Spectrum Length}
  
The Fourier analysis method was applied to two separate wavelength
ranges.   The two regions of spectra, 560.0--608.0 nm (red wavelength
range) and 440.0--515.0 nm (blue wavelength range), are similar in total wavelength span,  however, due to the constant bin width in logarithmic space, the blue region has many more pixels in total. 
A large number of lines are required to increase the planet signal above
the level of the noise \citep{Liu2007, Cameron1999}. For the Fourier analysis method, the blue wavelength range is more
sensitive to smaller planet signals (see Fig.~\ref{Fourierplots}) due to the increased density of sharp absorption lines that have a larger signal in the Fourier domain. 
For the application of the Collier Cameron method to the mock spectra,
with ideal stellar spectra removal, this is also true. However, for real
data, where the stellar spectrum is not exactly known, a high density of
absorption lines in the stellar spectrum is expected to increase the difficulty in fitting the continuum and removing the stellar spectrum without affecting the planet signal. 
For many methods, spectra with wavelengths towards the red are better
suited to detecting the planet signal, as absorption lines are scattered
into regions of the spectrum without strong stellar lines
\citep{Liu2007}.  The continuum regions in the red wavelength range also
aid the removal of the stellar spectrum in the Collier Cameron method, as the template spectrum can be more accurately scaled.

Using a shorter wavelength range is beneficial for decreasing the effect of assuming a wavelength-independent albedo. The Fourier analysis method shows gradual improvement in sensitivity to smaller planet-to-star fluxes for spectra covering a larger wavelength span, due to the increased number of absorption lines constraining the moving signal. In practice this is limited by the regions of bad pixels on a detector, as the Fourier analysis method requires continuous spectra. 

\subsubsection{Number of Exposures}    

For alternative methods of measuring the scattered light contribution of
a non-transiting planetary system such as the Collier Cameron method,
large numbers of spectra are required to fully cover the orbital phase
range of the planet and determine the planet's velocity \citep{Cameron2002}. This is not required for transiting systems, where the velocity of the planet is known for the full orbit. As the scattered light signal from the planet is intrinsically dim, taking spectra at maximum planetary phase increases the chances of measuring a planet flux. 

Thirty exposures were used to measure the planet-to-star flux of the ideal system in testing the Fourier analysis and the Collier Cameron method. This allows for a range of planet velocities, with differing Doppler shifts, but limits the processing time and the spread of planet phases.  The velocity must be large enough to shift the planet signal clear of the corresponding stellar features.  Using a shorter exposure would yield more images within the possible orbital phase range, but reduce the signal in each exposure. 

For up to 50 exposures in a phase range of 0.55--0.60 there is an improvement in the sensitivity to smaller planet-to-star fluxes of the Fourier analysis method, due to the stronger constraints on the moving signal. This results in exposures where the planet velocity differs by a couple of m/s in each exposure. 

The number of exposures with sufficiently different velocities is limited by the range within which the planet phase does not rapidly vary, and the length of the exposure required for an adequate  signal-to-noise ratio.
It is important that the separations in the planet's velocity from one observation to the next are not precisely uniform, as the solutions can blow up when the wavelength of the Fourier component is some multiple of the frequency spacing. 
The number of spectra used for the Monte Carlo simulations is consistent with the number of spectra available for the Subaru~HDS data of HD 209458, which we discuss in the following Section.

\subsection{Summary}

The Fourier analysis method
is generally more sensitive to smaller planet-to-star fluxes than the Collier
Cameron method for the ideal test spectra presented here, due to the ability to
extract the planet spectrum without having to remove a modelled version
of the stellar spectrum. 
At the $1\sigma$ level, the Fourier analysis and the Collier Cameron
method are similarly sensitive to small planet-to-star fluxes in the
blue wavelength range.
In the red wavelength range, the Fourier analysis
method is around three times more sensitive than the Collier Cameron method
at the $1\sigma$ level. At the $2$ and $3\sigma$ levels, the Fourier analysis method is twice as  
sensitive as the Collier Cameron method in the blue wavelength range. 
The Fourier analysis method
utilises the known parameters in the transiting planet case,
and allows a more sensitive measurement from a small portion of the orbit
that can be observed over a short amount of time. This is useful for
observations with ground-based telescopes for follow up observations to
space-based discoveries. 

The Fourier analysis method developed here is best suited to studying hot Jupiter planets with orbital periods of a few days, and spectra taken of bright host stars in the blue wavelength range, just before or after a secondary eclipse. 
Observations of transiting hot Jupiter planet's should be planned for when the planet is furthest from the observer and therefore appears close to full phase and maximum reflectance. The planet should have a velocity large enough to shift the scattered spectrum clear of the corresponding stellar absorption lines. 
When the wavelength of the Fourier component is some multiple of a
precisely uniform velocity spacing of the observations,  the matrices
become singular and the solutions can blow up. Therefore, observations
should be planned with some intentional irregularity in velocity
differences. 

Observations taken over a short velocity range are
beneficial to reduce the variation due to the unknown planet phase function. 
The relatively small number of exposures required to obtain a
measurement of the planet-to-star flux with the Fourier analysis method
is  therefore beneficial for rapid ground-based follow up to detections
by space-based surveys. 
The Fourier analysis method is also suited to measuring the planet-to-star flux over short wavelength ranges and therefore determining the likelihood of hot Jupiter planets having wavelength dependent albedos in the optical.

\section{Subaru HDS Data}

In this Section we use Spectra of HD~209458 and HD~189733 to test the Fourier analysis method for detecting light scattered by a transiting extrasolar planet with ground-based capabilities.
This data set includes 32 \'{e}chelle spectra from observations taken 2002 October 26 of HD~209458 with Subaru HDS with an $0.8''$ slit \citep{HDS} and  47 \'{e}chelle spectra from observations taken 2008 August 26 of HD~189733 with Subaru HDS with a $0.4''$ slit. 
The total wavelength ranges of the CCDs are 554--680~nm for the red, and 415--550~nm for the blue when the orders are combined.

The HD~209458 data has a spectral resolution of ${\textrm R}=\lambda/\Delta\lambda=45,000$, and a   signal-to-noise ratio per pixel of $\sim$350 for a 500 second exposure. For details of the observations of HD~209458b see \citet{Winn2004} and \citet{Narita2005}.  
The HD~189733 data has a spectral resolution of ${\textrm R}=\lambda/\Delta\lambda=90,000$ and a typical   signal-to-noise ratio per pixel of $\sim$300 for a 500 second exposure.
During the observations HD~209458b had just left a secondary eclipse and had velocities in the range 30--80~${\textrm k \textrm m \  \textrm  s^{-1}}$ and HD~189733b had velocities of 60--130~${\textrm  k \textrm m \ \textrm s^{-1}}$ just prior to a secondary eclipse. 

The \'{e}chelle spectra taken were processed using standard IRAF procedures, following a correction for the non-linearity of the HDS CCDs using the function measured by \
{Tajitsu2010}.
The one-dimensional spectra need to be normalised for variation in the efficiency of the CCD between the peak at the centre and the low count level close to the edge.
An $11$-knot cubic-spline was used to normalise the large scale variation of the response function of the red and blue CCDs, without removing the weak planet signal and the absorption lines.
The simplest method of combining the orders of each CCD by averaging the overlapping regions of the normalised spectra can magnify the noise and was therefore avoided.
Instead, the object spectra and the response functions were separately summed.
The combined spectra were then divided by the combined response function to remove large scale profiles. This preserves the signal in the overlapping regions. 
Bad pixels in the centre of the red CCD and at the edges of the blue CCD limit the full wavelength range possible for a continuous spectrum.

\subsection{Correction for Instrumental Variations}
\label{blaze_correction}

Before the combination of the dispersed orders, the variability of the spectra with time due to instrumental effects can be removed via a CCD response and wavelength correction as outlined in \citet{Winn2004}.  
The mean ratio of two spectra taken at different times should be equal to unity. This is not the case for the Subaru HDS data, as the ratio changes with wavelength bin. 
The spectrum format is affected by instrumental effects such as flexure of the spectrograph, changes in the temperature, changes in the inclination of the \'{e}chelle and cross-dispersion gratings and the flux calibration \citep{Aoki2002}. 

The temperature can change by around 0.1 degrees per hour, causing a 0.14 CCD pixel shift of the spectrum. Changes in the collimator mirror or the inclinations of the gratings can cause up to 1 pixel difference in the repeatability of the spectrograph set up. The largest variation within a night of observations comes from the flux calibration, which can vary the profile of the orders of the spectra in different exposures by up to 10~per cent \citep{Suzuki2003}. 

 The variation between exposures increases with time, such that the largest variation is seen between the first and last exposure. Following the process outlined in \citet{Winn2004}, the pattern of the variation of the ratio with wavelength bin can be used to correct adjacent orders to scale the flux calibration to match the first exposure. 
The ratio of the $n^{th}$ order of two spectra in the $j^{th}$ wavelength bin is given by;
\begin{equation}
{\textrm R}(\lambda_{n,j}) = \frac{\textrm S_{1}(\lambda_{n,j})}{\textrm S_{2}(\lambda_{n,j})},
\end{equation}
where R signifies the ratio, $\textrm S_{1}$ is the first exposure and $\textrm S_{2}$ is a later exposure~\citep{Winn2004}.
The ratios of the exposures for orders adjacent to the one being corrected are boxcar smoothed over $j$, with a width of 100 pixels. 
The flux corrected $n^{th}$ order of the second spectrum ${\textrm S_{2 \textrm b}(\lambda_{n})}$ is given by;
\begin{equation}
{\textrm S_{2 \textrm b}}(\lambda_{n,j}) = [{\textrm c_{n+1}}{\textrm R'}(\lambda_{n+1,j}) + {\textrm c_{n-1}}{\textrm R'}(\lambda_{n-1,j})]{\textrm S_{2}}(\lambda_{n,j}),
\end{equation}
where ${\textrm R'}(\lambda_{n})$ signifies the boxcar smoothed ratio of the $n^{th}$ order of the spectra \citep{Winn2004}.
Linear regression is used to find the factors ${\textrm c_{n+1}}$ and ${\textrm c_{n-1}}$ such that the sum of squared residuals is a minimum between ${\textrm S_{1}(\lambda_{n})}$ and the corrected second spectrum.

A correction can also be made for small variations in wavelength.  These variations are of the order of a pixel, corresponding to a velocity shift of less than $2~{\textrm k \textrm m \textrm s^{-1}}$, compared to an average planet velocity of $70~{\textrm k \textrm m \textrm s^{-1}}$.     The $n^{th}$ order of the matched spectrum  ${\textrm S_{2 \textrm m}(\lambda_{n})}$ is given by;
\begin{equation}
{\textrm S_{2 \textrm m}}(\lambda_{n,j}) = {\textrm c_{o}}{\textrm S_{2 \textrm b}}(\lambda_{n,j}) + \Delta \lambda \frac{\textrm d \textrm S_{2 \textrm b}(\lambda)}{\textrm d \lambda},
\end{equation}
where the derivative is determined via 3-point Lagrangian interpolation \citep{Winn2004}. 
Linear regression is used to find the factors ${\textrm c_{0}}$ and $\Delta \lambda$ such that the sum of squared residuals is a minimum between ${\textrm S_{1}(\lambda_{n})}$ and the matched second spectrum. 

For the Subaru~HDS data, orders at wavelengths shorter than the location of the bad pixels in the centre of the red CCD were corrected, corresponding to a range of 558.83--599.56~nm. The blue wavelength range of 445.28--508.11~nm was also corrected before normalisation and combination.

\begin{table*}
\centering
\begin{minipage}{140mm}
\begin{tabular}{c c c c}
\hline
&& HD 209458 system & HD 189733 system \\ [0.1cm]
\hline
 {\scriptsize Orbital Period} & $ {\textrm P_{\textrm o \textrm r \textrm b}}$  &$3.52474859\pm0.00000038$  {\scriptsize days}   & $2.218573\pm0.00002$  {\scriptsize days}  \\ [0.1cm]
{\scriptsize Orbital Inclination} &  {\it i} & $86.929^{+0.009}_{-0.010}$  {\scriptsize degrees} &  $85.79\pm0.24$  {\scriptsize degrees}\\ [0.1cm]
 {\scriptsize  Mass of Planet b} &  ${\textrm M_{\textrm p}}$ & $0.64\pm0.06$  ${\textrm M_{\textrm J}}$ & $1.154\pm0.033$  ${\textrm M_{\textrm J}}$\\ [0.1cm]
{\scriptsize Mass of Host Star} & ${\textrm M_{\star}}$ & $1.101^{+0.066}_{-0.062}$  ${\textrm M_{\odot}}$ &  $0.82\pm0.03$  ${\textrm M_{\odot}}$ \\ [0.1cm]
 {\scriptsize Transit Epoch} & ${\textrm T_{0}}$ & $2,452,826.628521\pm0.000087$  {\scriptsize days} & $2,453,988.80336\pm0.00024$  {\scriptsize days} \\ [0.1cm]
 {\scriptsize Semi-Major Axis} & a & $0.04704\pm^{0.00048}_{0.00047}$  {\scriptsize AU} & $0.03099\pm^{0.0006}_{0.00063}$  {\scriptsize AU} \\ [0.1cm]
\hline
\end{tabular}
 \caption[Physical parameters of HD 209458b and HD 189733b]{Physical parameters of the star HD 209458 with transiting planet HD 209458b \citep{Knutson2007} and of the star HD 189733 with transiting planet HD 189733b \citep{Bouchy2005,Bakos2006}.  \label{HDtable}}
\end{minipage}
\end{table*}

\section{Results}

In order to predict the wavelength shift of the scattered spectra with respect to the stationary stellar spectra, the velocity of the planet was determined from the physical parameters listed in Table \ref{HDtable} and the Julian date at the time of each exposure. 
During the observations HD~209458b had just left a secondary eclipse and had velocities toward the observer in the range 30--80~${\textrm k \textrm m \  \textrm  s^{-1}}$ and an average planet phase of 0.88.   Just prior to a secondary eclipse HD~189733b had velocities of 60--130~${\textrm  k \textrm m \ \textrm s^{-1}}$ away from the observer and an average planet phase of 0.74.

The expected level of sensitivity of the Fourier analysis method can be determined for the observational parameters and the Monte Carlo simulations presented in Section~\ref{jenkins}. 
Fig.~\ref{Fourierplots} shows contours of the level of detection of
planet-to-star fluxes in the range $10^{-5}$--$2\times10^{-4}$ and
signal-to-noise ratio values of $10^{2}$--$10^{3}$ for the red and the
blue wavelength ranges. The dot-dashed line indicates the observational   signal-to-noise ratio for the Subaru HDS \'{e}chelle spectra of HD~209458 of 350. The upper limit of $\epsilon < 1.6\times10^{-5}$ set by \citet{Rowe2008} with the MOST satellite is indicated by the dashed line. 
Upper limits can be derived from non-detections based on these Monte Carlo results, however the previous upper limit for HD~209458b is below the $1\sigma$ level of detection for this data with the Fourier analysis method. 

The spectra were analysed in the red (558.83--599.56~nm) and blue (445.28--508.11~nm) wavelength ranges separately, as the Fourier analysis method requires continuous spectra.  For the red and blue wavelength ranges of the HD~209458 data, the solutions were $\epsilon=7.45\times10^{-4}$ and $\epsilon={\textrm-}1.50\times10^{-4}$ respectively.  
A negative planet-to-star flux arising from the noise, as seen in the blue wavelength range solution, is not physical but may be used to statistically constrain the upper limit on the albedo. 
The red result is larger than expected by theory and the upper limit on the planet-to-star flux of HD~209458b of $\epsilon<1.6\times10^{-5}$ set by \citet{Rowe2008}.  
Similarly, for the red  and blue wavelength ranges of the HD~189733 data, the solutions were $\epsilon=2.45\times10^{-4}$ and $\epsilon={\textrm-}1.08\times10^{-4}$ respectively. \\

\begin{figure}
\begin{center}
  \includegraphics[width=80mm]{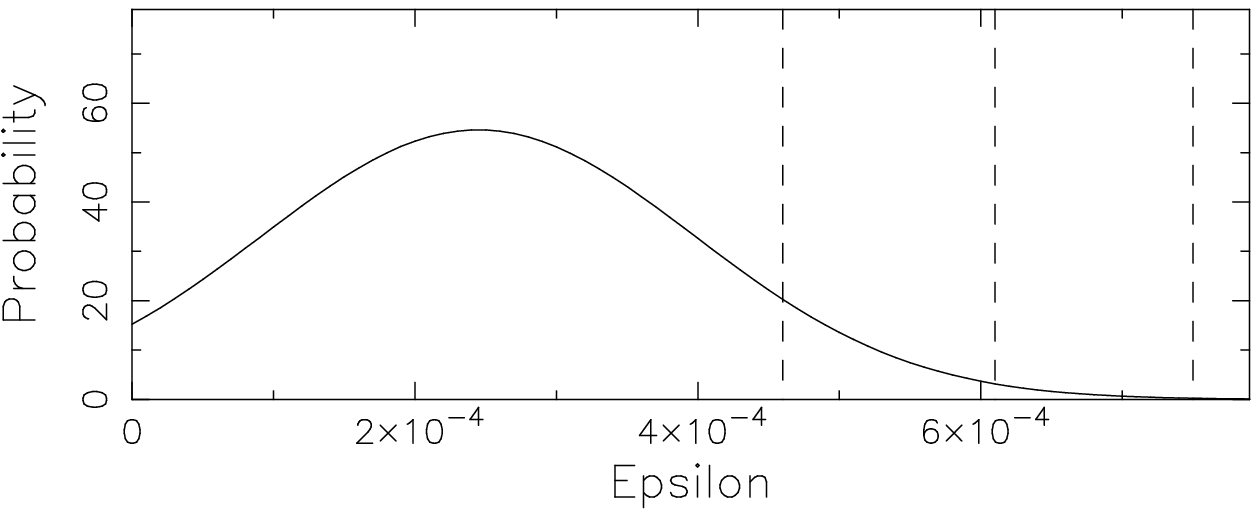}
 \caption[Probability curve for HD~189733]{{\bf Probability curve for HD~189733 in the red wavelength range.} ${\textrm P}(\epsilon_{\textrm t\textrm r\textrm u\textrm e}|\epsilon_{\textrm s\textrm p\textrm e\textrm c\textrm t\textrm r\textrm a})$, normalised to an area of unity. The dashed lines show the $1\sigma$, $2\sigma$ and $3\sigma$ upper limits on the planet-to-star flux given the results of the analysis of the HD~189733 spectra. These correspond to upper limits of $\epsilon<4.6\times10^{-4}$, $\epsilon<6.1\times10^{-4}$ and $\epsilon<7.5\times10^{-4}$ respectively in the wavelength range 558.83--599.56~nm. \label{HD189733_prob}}
\end{center} 
\end{figure} 

An upper limit can be set on the planet-to-star flux via probability distributions from Monte Carlo simulations and the physical constraint that the actual planet-to-star flux must be positive. From  Bayes' theorem we have:
\begin{equation}
{\textrm P}(\epsilon_{\textrm t\textrm r\textrm u\textrm e}|\epsilon_{\textrm s\textrm p\textrm e\textrm c\textrm t\textrm r\textrm a}) \propto {\textrm P}(\epsilon_{\textrm s\textrm p\textrm e\textrm c\textrm t\textrm r\textrm a}|\epsilon_{\textrm t\textrm r\textrm u\textrm e}){\textrm P}(\epsilon_{\textrm t\textrm r\textrm u\textrm e}),
\end{equation}
where ${\textrm P}(\epsilon_{\textrm t\textrm r\textrm u\textrm e}|\epsilon_{\textrm s\textrm p\textrm e\textrm c\textrm t\textrm r\textrm a})$ is the probability that the measured planet-to-star flux is the physically correct value given the data. ${\textrm P}(\epsilon_{\textrm s\textrm p\textrm e\textrm c\textrm t\textrm r\textrm a}|\epsilon_{\textrm t\textrm r\textrm u\textrm e})$ is the likelihood that the data will result in the measured planet-to-star flux. The prior ${\textrm P}(\epsilon_{\textrm t\textrm r\textrm u\textrm e})$  constrains the physical planet-to-star flux to be positive.

Monte Carlo simulations were run for the simulated data with the exact observational parameters of the Subaru~HDS data in order to determine ${\textrm P}(\epsilon_{\textrm s\textrm p\textrm e\textrm c\textrm t\textrm r\textrm a}|\epsilon_{\textrm t\textrm r\textrm u\textrm e})$, the likelihood that the data will result in the measured planet-to-star flux.
For the HD~209458 results, 32 spectra were generated with a   signal-to-noise ratio of 350, a spectral resolution of 45,000 and an orbital phase range of 0.55--0.60. For the HD~189733 results, 25 spectra were generated with a   signal-to-noise ratio of 300, a spectral resolution of 90,000 and an orbital phase range of 0.34--0.44.

The likelihood of the results of the HD~209458 and HD~189733 data in the blue wavelength range shows a very small probability of having a positive and therefore physical planet-to-star flux. This suggests that the data correction did not fully account for the time varying signal due to the high density of absorption lines in the blue wavelength range, and the results measured with the Fourier analysis method are spurious. Similarly, the results for the red wavelength range of the HD~209458 spectra are thought to be spurious as the likelihood shows a very small probability of having a physical planet-to-star flux below the previously defined upper limit of $\epsilon<1.6\times10^{-5}$ set by \citet{Rowe2008}. 

Fig.~\ref{HD189733_prob} shows the results of the Bayesian analysis for the red wavelength range of the HD~189733 data. The measured planet-to-star flux was $\epsilon = 2.45\times10^{-4}$. The likelihood from the Monte Carlo simulations was fit with a Gaussian distribution and multiplied by the prior condition that the planet-to-star flux must be positive.
The probability that the measured planet-to-star flux is the physically correct value given the data, ${\textrm P}(\epsilon_{\textrm t\textrm r\textrm u\textrm e}|\epsilon_{\textrm s\textrm p\textrm e\textrm c\textrm t\textrm r\textrm a})$,  is then normalised to have an area of unity. From this curve the $1\sigma$, $2\sigma$ and $3\sigma$ upper limits on the planet-to-star flux of HD~189733b in the wavelength range 558.83--599.56~nm are found to be $\epsilon<4.6\times10^{-4}$, $\epsilon<6.1\times10^{-4}$ and $\epsilon<7.5\times10^{-4}$ respectively. 

\section{Analysis}

To investigate the level of spurious signals in the data we  shuffled Doppler shifts with respect to the corresponding exposures. This was performed for both stars in both wavelength ranges. A distribution of planet-to-star fluxes was measured due to spurious time variable signals arising out of the instrumental noise. 
Monte Carlo simulations were then used to determine the distribution of false planet-to-star fluxes that are not correlated with the Doppler shifts of the planet signal.
Random velocities were selected in the same range as the real data; in the range 30--80~${\textrm k \textrm m \  \textrm  s^{-1}}$ for HD~209548 and 60--130~${\textrm  k \textrm m \ \textrm s^{-1}}$ for HD~189733. This was also done for the mock data.  

\begin{figure}
\begin{center}
  \includegraphics[width=80mm]{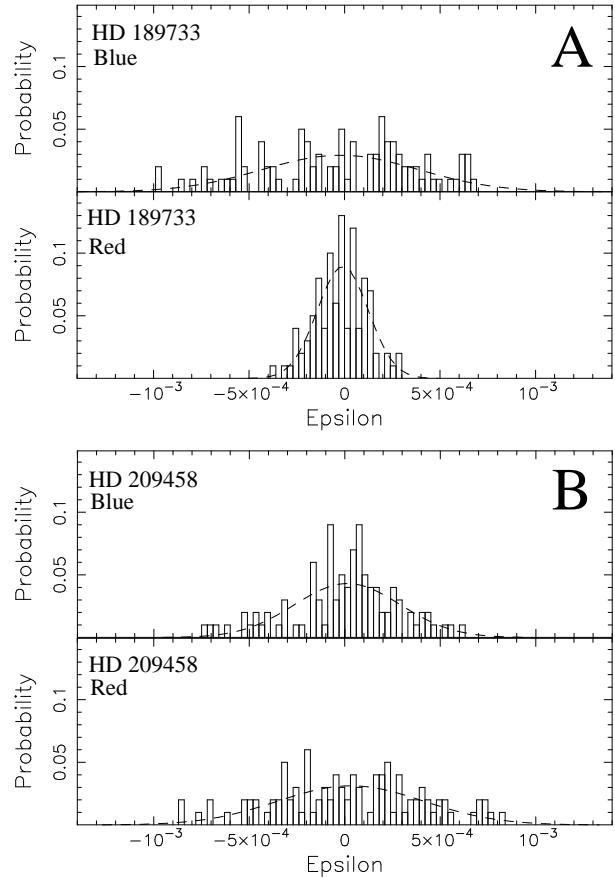}
 \caption[Probability curves for Subaru data]{{\bf Probability curves.} Histograms from Monte Carlo simulations showing the probability of measuring an $\epsilon_{\textrm m \textrm e \textrm a \textrm s \textrm u \textrm r \textrm e \textrm d}$ with shuffled planet velocities used to obtain the solutions. Panel {\bf A} - Results for the HD~189733 Subaru data. Panel {\bf B} - Results for the HD~209458 Subaru data. {\it Upper subpanels} - Results for data in the blue wavelength range. {\it Lower subpanels} -  Results for data in the red wavelength range. The dashed curves show the Gaussian distribution with the same mean and standard deviation. \label{probs}}
\end{center} 
\end{figure}

\begin{figure}
\begin{center}
  \includegraphics[width=80mm]{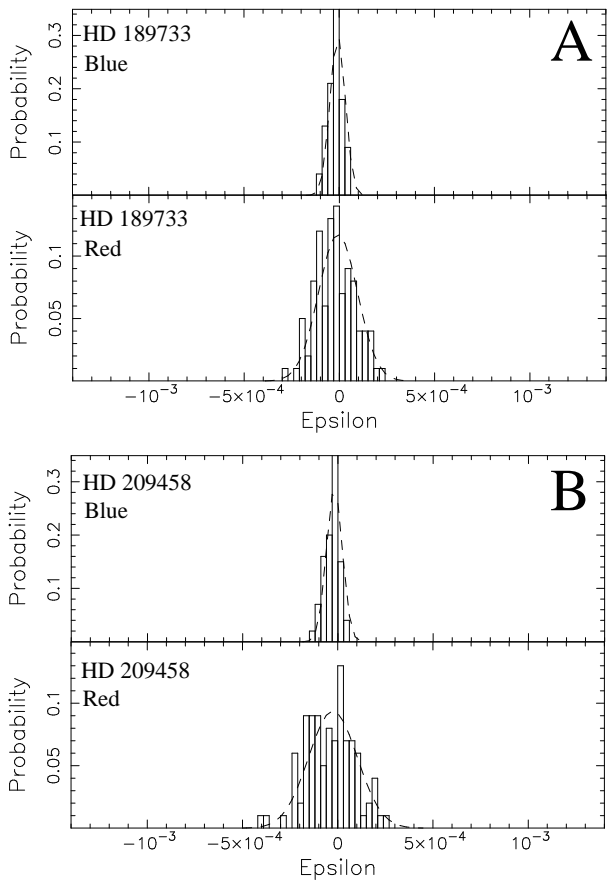}
 \caption[Probability curves for mock spectra]{{\bf Probability curves.} Histograms from Monte Carlo simulations showing the probability of measuring an $\epsilon_{\textrm m \textrm e \textrm a \textrm s \textrm u \textrm r \textrm e \textrm d}$ with shuffled planet velocities used to obtain the solutions. Panel {\bf A} - Results for the HD~189733 mock data. Panel {\bf B} - Results for the HD~209458 mock data. {\it Upper subpanels} - Results for data in the blue wavelength range. {\it Lower subpanels} -  Results for data in the red wavelength range. The dashed curves show the Gaussian distribution with the same mean and standard deviation. \label{probs_test}}
\end{center} 
\end{figure}

The distribution of $\epsilon_{\textrm m \textrm e \textrm a \textrm s \textrm u \textrm r \textrm e \textrm d}$ values for randomly selected Doppler shifts are compared for both wavelength ranges of the HD~209458 and HD~189733 Subaru HDS spectra in Fig.~\ref{probs}. The distributions for the corresponding simulated mock spectra are shown in Fig.~\ref{probs_test}. 
Although the data corrections for instrumental effects reduce the spread of the spurious signals for the Subaru~HDS spectra, it is apparent that the time variability of the spectrum has not been fully removed. This can be seen by the comparison of the spread of the false-planet signals for the mock data, compared to the corrected Subaru~HDS data.

Similar data corrections based on low order polynomial corrections to the CCD response and the cross-correlation of spectral lines have been adopted for spectra taken with Subaru~HDS \citep{Narita2005, Snellen2008}, the High-Resolution Echelle Spectrograph (HIRES) on Keck \citep{Suzuki2003} and the High Resolution Spectrograph (HRS) on the Hobby-Eberly Telescope (HET) \citep{Redfield2008}. 

\citet{Narita2005} found the correction in \citet{Winn2004} to be accurate to approximately the Poisson noise limit for the HD~209458 data analysed here, and that the main challenge in using ground-based spectra for high precision measurements is instrumental variation, not the removable telluric contribution. 
\citet{Suzuki2003} calibrate the flux of Keck HIRES spectra to within 1~per cent, correcting for vignetting and extraction using reference spectra. 
This suggests that instruments with increased stability are required for more accurate ground-based follow up characterisation of extrasolar planets. 

The flux and wavelength calibration correction reduced the spread of the spurious signals seen in the distributions in Fig.~\ref{probs} by an order of magnitude.  In the mock data the blue wavelength ranges have narrower distributions of false planet signals, suggesting that the increased density of absorption lines in this region is more likely to constrain the moving signal to be correlated with the planet velocity. This is not seen in the HD~189733 corrected data. This may be partly due to the $11$-knot cubic-spline chosen to remove the shape of the response of the CCDs being less suitable for spectra with a high density of lines. Additionally,  in practice, the wavelength scale and flux calibration correction may be  harder to implement for the higher density of spectral lines in the blue wavelength range. 

The red wavelength range of the corrected HD~189733 spectra is the closest to the mock spectra. It can be seen in Fig.~\ref{probs} and Fig.~\ref{probs_test} that the two distributions of false-planets have a similar spread. Therefore the time variable signal due to instrumental effects has been satisfactorily removed, and the corrected data is comparable to the ideal simulated data. From this corrected data the $1\sigma$ upper limit on the planet-to-star flux in the wavelength range 558.83--599.56~nm is $\epsilon<4.6\times10^{-4}$. 
This upper limit is not sensitive enough to constrain atmospheric models of hot Jupiter planets or rule out a highly reflective upper cloud deck. 

These instrumental variations would also hinder the application of the Collier Cameron method to this Subaru HDS data. It is beyond the scope of this paper to apply the Collier Cameron method to the non-idealised spectra. Future searches for the light scattered by an extrasolar planet will continue to adopt the Collier Cameron method, and the method developed by \citet{Charbonneau1999}. These methods are still required for non-transiting planets. For transiting planets, the application of these methods with constrained parameters may yield more stringent upper limits.

The red wavelength HD~189733 data set showed successful implementation of instrumental corrections to Subaru~HDS spectra and the application of the Fourier analysis method. The archive data obtained to test the Fourier analysis method did not have sufficient   signal-to-noise ratio to significantly constrain the planet-to-star flux of the systems. Improved stability of the instrument while observing bright systems with high   signal-to-noise ratio will be required for follow up characterisation of transiting systems detected by space-based missions via the Fourier analysis method.

\section{Conclusion}

Measuring planet-to-star flux ratios will continue to guide theories on the composition of the upper atmosphere of hot Jupiter planets and methods of radiation transfer. This is complementary to the detections of thermal emission from transiting very hot Jupiter planets in the optical and infrared. However, previous upper limits set by non-detections of the light scattered by systems such as  $\tau$~Bo\"{o}tis \citep{Charbonneau1999, Cameron1999, Leigh2003a} and HD~75289A \citep{Rodler2008,Leigh2003b} support the need for  model independent approaches that exploit the benefits of the numerous transiting systems to minimse the unknown orbital parameters. It is also advantageous to measure phase independent planet-to-star fluxes without assuming a grey albedo. 

In this paper we have introduced a new model-independent method for isolating the scattered starlight signal from the host star flux in high resolution spectra, that is suited to typical transit data.  
Using mock spectra this Fourier analysis method was shown to be more appropriate for typical observations of a well constrained transiting system, and therefore more sensitive to smaller planet-to-star fluxes, than the currently used Collier Cameron method.
Using Monte Carlo simulations the sensitivity to dim planets of the Fourier analysis method was shown to increase with higher   signal-to-noise ratio data, and to be better suited to the blue region of the visible spectrum due to the increased signal in Fourier space of the numerous deep absorption lines. 
The Fourier analysis method is more likely to detect the planet-to-star flux of a dim planet such as HD~209458b, with  Subaru~HDS and  ideal high   signal-to-noise ratio spectra, than the Collier Cameron method. 

The Fourier analysis method for extracting the light scattered by transiting hot Jupiter planets from high resolution spectra was applied to \'{e}chelle spectra of HD~209458 and HD~189733. Due to instrumental variations that could not be fully corrected for, there was no improvement on the measurement of the upper limit of the planet-to-star flux of HD~209458 compared to the current limit set by \citet{Rowe2008}. A $1\sigma$ upper limit on the planet-to-star flux of HD~189733b was measured in the wavelength range of 558.83--599.56~nm of $\epsilon<4.5\times10^{-4}$. This result does not constrain atmospheric models. A deeper measurement of the upper limit of planet-to-star flux of this system with ground-based capabilities requires data with a higher   signal-to-noise ratio, and increased stability of the telescope. 

Radial velocity surveys observe stars in a brighter magnitude range than transiting photometry surveys \citep{Borucki2003,Tinney2001,Udry2000}. For example Kepler is surveying stars of magnitude 14--9, as compared to the Anglo-Australian Planet Search (AAPS) on the Anglo-Australian Telescope (AAT) which targets stars brighter than V$<7.5$. Therefore measuring the light scattered by the upper atmosphere of planets will most likely require candidates selected via radial velocity methods and followed up with transit photometry, to ensure that adequate   signal-to-noise ratio is obtained. 

The next generation of extremely large telescopes (ELTs) including the European Extremely Large Telescope (E-ELT), the Giant Magellen Telescope (GMT) and the Thirty Meter Telescope (TMT), will be the largest optical and infrared telescopes ever built, dramatically increasing the sensitivity of current ground-based telescopes \citep{Carlberg2005}. The reflecting mirror diameters range from 25--42~m.  With a diameter of 40~m, the collecting area and hence the signal-to-noise will be increased by a factor of $\sim$5 compared to similar observations with current 8~m ground-based telescopes. This will enable higher signal-to-noise ratios to be obtained \citep{Gilmozzi2007}. Therefore, smaller planet-to-star fluxes of transiting hot Jupiter planets will be measurable with ELTs and the Fourier analysis method.

\section*{Acknowledgments}

NSO/Kitt Peak FTS data used here were produced by National Optical Astronomy Observatory/Association of Universities for Research in Astronomy/National Science Foundation

This paper is partly based on data collected at Subaru Telescope, which is operated by the National Astronomical Observatory of Japan. 

We thank the anonymous referee for their detailed comments, helping us to improve the paper.

S.V.L aknowledges the support of an Australian Postgraduate Award, an Albert Shimmins Postgraduate Writing-Up Award, a
Postgraduate Overseas Research Experience Scholarship, a Melbourne Abroad Traveling Scholarship and the hospitality of Princeton
University Observatory during two extended visits in 2007 and 2008.

N.N. is supported by a Japan Society for Promotion of Science (JSPS) Fellowship for Research (PD: 20-8141).

Y.S. acknowledges the support from the Global Scholars Program of Princeton University. This work is partially supported by Grant-in-Aid for Scientific research of Japanese Ministry of Education, Culture, Sports, Science and Technology (No.20340041), and by JSPS Core-to-Core Program ``International Research Network for Dark Energy''.

\bibliography{mybib}
\bibliographystyle{mn2e}

\bsp

\label{lastpage}

\end{document}